\documentclass{article}
\usepackage{honnef,pslatex,epsfig,wrapfig}

\newcommand{\cray} {{\sc cr}}
\newcommand{\eas} {{\sc eas}}
\newcommand{\gzk} {{\sc gzk}}

\frompage{000} \topage{000}

\def\la{\mathrel{\mathchoice {\vcenter{\offinterlineskip\halign{\hfil
$\displaystyle##$\hfil\cr<\cr\sim\cr}}}
{\vcenter{\offinterlineskip\halign{\hfil$\textstyle##$\hfil\cr<\cr\sim\cr}}}
{\vcenter{\offinterlineskip\halign{\hfil$\scriptstyle##$\hfil\cr<\cr\sim\cr}}}
{\vcenter{\offinterlineskip\halign{\hfil$\scriptscriptstyle##$\hfil\cr<\cr
\sim\cr}}}}}

\def\ga{\mathrel{\mathchoice {\vcenter{\offinterlineskip\halign{\hfil
$\displaystyle##$\hfil\cr>\cr\sim\cr}}}
{\vcenter{\offinterlineskip\halign{\hfil$\textstyle##$\hfil\cr>\cr\sim\cr}}}
{\vcenter{\offinterlineskip\halign{\hfil$\scriptstyle##$\hfil\cr>\cr\sim\cr}}}
{\vcenter{\offinterlineskip\halign{\hfil$\scriptscriptstyle##$\hfil\cr>\cr
\sim\cr}}}}}

\def\rightmark{Cosmic Rays and Particle Physics} 
\def\leftmark{K.-H. Kampert}
 
\title{Cosmic Rays and Particle Physics} 
\authors{{Karl-Heinz Kampert}\\[2.812mm]
{\normalsize
Universit\"at Karlsruhe (TH), Institut f\"ur Experimentelle 
Kernphysik\\[0.2ex] 
Forschungszentrum Karlsruhe, Institut f\"ur Kernphysik\\
P.O.B. 3640, D-76021 Karlsruhe, Germany
}}
 
\abstract{The study of high energy cosmic rays is a diversified
field of observational and phenomenological physics addressing
questions ranging from shock acceleration of charged particles in
various astrophysical objects, via transport properties through
galactic and extragalactic space, to questions of dark matter, and
even to those of particle physics beyond the Standard Model
including processes taking place in the earliest moments of our
Universe.  After decades of mostly independent evolution of
nuclear-, particle- and high energy cosmic ray physics we find
ourselves entering a symbiotic era of these fields of research. 
Some examples of interrelations will be given from the
perspective of modern Particle-Astrophysics and new major
experiments will briefly be sketched.}
 
\begin{document}
 
\maketitle
\vspace*{24pt}

\section{Introduction}

Cosmic rays (\cray s) were discovered in 1911 by Victor Hess
through a series of balloon flights in which he carried
electrometers to over 5000 m \cite{hess-1912}.  Originally being
thought of as penetrating $\gamma$-radiation, in the late
twenties Compton and others realized that \cray s mainly consist
of charged particles.  By performing coincidence measurements in
1938 using Geiger counters at mountain altitudes and later also
at sea level in Paris, Pierre Auger discovered the phenomenon of
``extensive air showers''; A high energy \cray\ entering the
atmosphere initiates a cascade of secondary particles which is
large enough and sufficiently penetrating to reach ground level. 
From his observations, Pierre Auger already concluded that
primary particles up to energies of $10^{15}$ eV are found in
\cray s, and speculations were raised how to generate particles
of such high energy.  Present day simulations predict that, e.g.\
a single $10^{15}$ eV \cray\ particle produces about $10^{6}$
secondary particles at sea level, mainly photons and electrons
plus some muons and hadrons being spread out over about a
hectare.  Indeed, the present particle physics has taken origin
from the observations and measurements of \cray s performed in
the first half of our century, starting from the discovery of the
positron in 1932, muons in 1937, to that of pions and strange
particles ($\Lambda$ and $K$) in 1947, $\Xi^-$ and $\Sigma^{+}$
in 1952-53, and possibly even to the discovery of charm in 1971
\cite{niu71}.  On the other hand, nuclear and particle physics
have also provided important input to \cray\ physics.  For
example, data of nuclear spallation cross sections measured at
accelerators turned out to be a key for understanding the
propagation of \cray s in our galaxy.  Also, phenomenological
prescriptions of high energy p+p, p+A and A+A interactions and
the modelling of a Quark-Gluon Plasma state enter directly into
Monte Carlo simulations of extensive air showers.  Now, when the
physics of accelerators is starting to fight against both
technological and financial limitations, we see that new interest
is flowing back to the origins \cite{richter00}.  In fact, many
fundamental and unresolved questions are still presented to us by
the cosmic radiation, and this is particularly true for the
extremely high energies.

\begin{wrapfigure}[29]{l}{8cm}
\epsfig{file=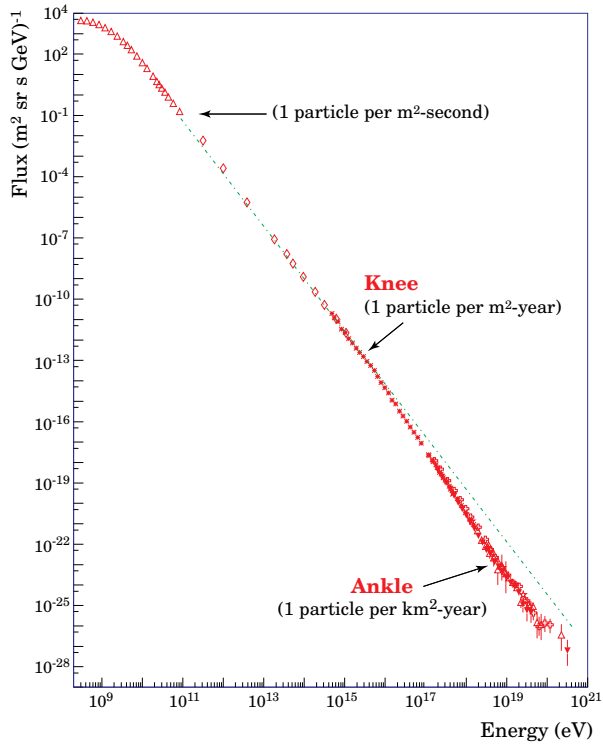,width=8cm}\vspace*{-4mm}
\caption{The cosmic ray all particle spectrum (adapted
from Ref.\ \cite{swordy-cr_spec}). Approximate
integral fluxes are indicated.}
\label{fig:cr-spectrum}
\end{wrapfigure} 
\noindent We know that the \cray\ energy spectrum extends from
below 1 GeV to above $10^{20}$ eV. The bulk of \cray s up to at
least an energy of some PeV ($10^{15}$ eV) is believed to
originate within our galaxy.  Above that energy, which is
associated with the so called ``knee'', the differential energy
spectrum of particles steepens from a power law $E^{-2.7}$ to
about $E^{-3.2}$.  Above the so called ``ankle'' at $E \simeq 5
\cdot 10^{18}$ eV, the spectrum flattens again to about
$E^{-2.8}$.  This feature is often interpreted as the cross-over
from a steeper galactic to a harder extra-galactic component. 
Figure \ref{fig:cr-spectrum} shows the measured \cray\ spectrum.

Up to energies of some $10^{14}$~eV the flux of particles is
sufficiently high so that their elemental distributions can be
studied by high flying balloon or satellite experiments.  Such
measurements have provided important implications for the origin
and transport properties of \cray s in the interstellar medium. 
Two prominent examples are ratios of secondary to primary
elements, such as the B/C-ratio, which are used to extract the
average amount of matter \cray-particles have traversed from
their sources to the solar system (5 - 10 g/cm$^{2}$), or are
radioactive isotopes, e.g.\ $^{10}$Be or $^{26}$Al, which carry
information about the average `age' of cosmic rays (1 - $2 \cdot
10^{7}$ a).

Above a few times $10^{15}$~eV the flux has dropped to only one
particle per m$^{2}$ and year.  This excludes any type of `direct
observation' even in the near future, at least if high statistics
is required.  Ironically, one of the most prominent features of
the \cray\ energy spectrum, the knee, is at an energy just above
some $10^{15}$~eV. It was observed already in 1956
\cite{kulikov56} but it still remains unclear as to what is the
cause of this spectral steepening.  In the standard model of
\cray\ acceleration the knee is attributed to the maximum energy
of galactic accelerators mostly believed to be supernova remnants
in the Sedov phase.

The other target of great interest is the energy range around the
Greisen-Zatsepin-Kuzmin ({\sc gzk}) effect at $E \simeq 5\cdot
10^{19}$ eV. Data currently exist, though with very poor
statistics, up to $3 \cdot 10^{20}$~eV and there seems to be no
end to the energy spectrum \cite{agasa00,nagano00b}.  Explanation
of these particles requires the existence of extremely powerful
sources within a distance of approximately 50-100 Mpc.  Hot spots
of radio galaxy lobes -- if close enough -- or topological
defects from early epochs of the universe would be potential
candidates.

Obviously, the topic of cosmic rays is very wide and deeply
related to many fields of physics, ranging from hydrodynamics and
astronomy via nuclear- and elementary particle physics to
questions of cosmology.  Experimentally, the topics addressed by
cosmic rays are closely related to TeV $\gamma$- and
$\nu$-astronomy, and to some aspects of dark matter searches, all
of which became known as ``Particle-Astrophysics''.  Since the
limited length of this paper excludes giving a comprehensive
review about all of these many interesting facets, we shall pick
only a few examples from different cosmic ray energy ranges and
discuss some experimental aspects.

\section{The Question of Dark- and Antimatter}

The questions for dark matter as a major contributor to the
energy density of the Universe or the Universe being a
patchwork consisting of distinct regions of matter and antimatter
are among the most fundamental ones in cosmology.  Both of them
are related to \cray\ measurements in different regions of energy.
In this section, we will discuss examples of direct observations 
on balloons and satellites.

\begin{wrapfigure}[28]{l}{8cm}
\vspace*{-4mm}\epsfig{file=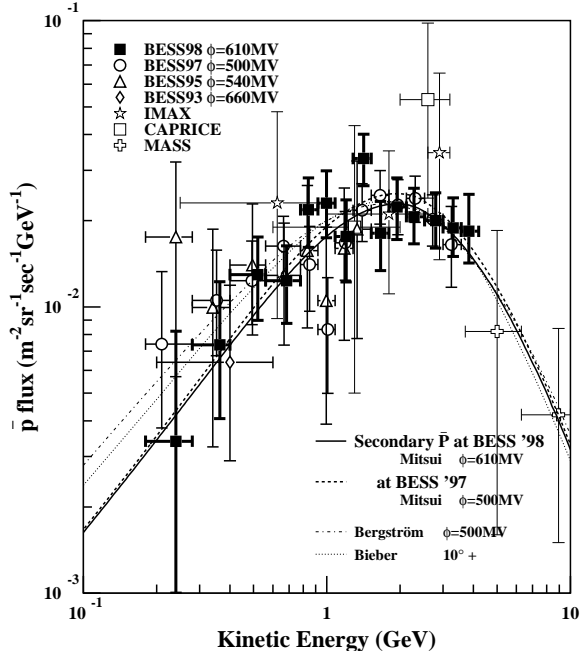,width=8cm}\vspace*{-8mm}
\caption{The {\sc bess} 1998 antiproton fluxes measured at the top of
the atmosphere.  The thick solid curve represents the expected
spectrum for secondary $\bar{p}$ at the 1998 flight.  Also shown are
other  previously existing data and calculations
\cite{bergstroem99,bieber99} for secondary $\bar{p}$ at solar minimum.
From Ref.\ \cite{maeno00}.}
\label{fig:pbar-spec}
\end{wrapfigure}

A representative detector of this type is the Japanese {\sc bess}
spectrometer.  The main parameters and components of this
detector are the 1 Tesla magnetic field produced by a thin (4
g/cm$^2$) superconducting coil filling a tracking volume equipped
with drift chambers providing up to 28 hits per track with an
acceptance of 0.3 m$^2$ sr.  In addition, two hodoscopes provide
d$E$/d$x$ and time-of-flight measurements.  A series of flights
performed between 1993 and 1998 yielded at total 848 $\bar{p}$'s
in the energy range 0.18 - 4.2 GeV. Their energy spectrum is
shown in figure~\ref{fig:pbar-spec} \cite{maeno00}.  The observed
peak around 2 GeV is a generic feature of secondary $\bar{p}$'s
which are produced by the interaction of galactic high energy
cosmic rays with the interstellar medium.  Both the shape of the
energy spectrum and the absolute flux is well reproduced by
several theoretical calculations \cite{simon98,bieber99}
using $\bar{p}$-production cross sections from nuclear physics
experiments \cite{gaisser99}.  However, there remains some
diversity about the calculations in the low energy region.  The
indication of an excess of antiprotons below 0.5 GeV has received
growing attention since its first observation by {\sc bess} in
1993.  Possible sources are the annihilation of neutralinos at
the galactic centre or the evaporation of primordial black holes
({\sc pbh}).  The latter may have been formed with arbitrarily small
masses during virulent conditions in the early Universe
\cite{novikov98}, e.g., by the collapse of large density
perturbations \cite{zeldovich66,hawking75}.  Data constraining
{\sc pbh} abundances will thus yield constraints on the density
fluctuation spectrum in the early Universe, an important
ingredient to structure formation theories.  For $M_{PBH} \la 4
\cdot 10^{13}$ g (typical mass of mountain) the evaporation process
will result in relativistic quarks and gluons which may produce
antiprotons during hadronisation.  The expectation for the flux
of antiprotons from this process is a spectrum increasing towards
lower kinetic energies down to $\sim$0.2 GeV \cite{macgibbon91}. 
Another strategy to look for dark matter particles is via
annihilation of neutralinos at the galactic centre;
$\chi\chi \to \ell^{+}\ell^{-},q\bar{q} \to \bar{p}, e^{+},
\gamma, \nu$.  The $\bar{p}$ flux from this source is
characterized by a significant flux below 1 GeV
\cite{bergstroem99}.  The uncertainty of the expected secondary
flux due to the uncertainty of \cray\ propagation in the Galaxy,
however, is at present still of the same order of magnitude as
the signal expected from these `primary' sources.

\noindent Measurements of antiparticles are directly linked also
to the search for primordial antimatter.  The laws of physics
treat matter and antimatter almost symmetrically, and yet the
stars, dust and gas in our celestial neighbourhood consist
exclusively of matter.  The absence of annihilation radiation
from the Virgo cluster shows that little antimatter is found
within typical sizes of galactic clusters and many cosmologists
assume that the local dominance of matter persists throughout the
entire visible universe.  However, observational evidence for a
universal baryon asymmetry is weak. 

As most of the $\bar{p}$'s originate from \cray\ interactions
with the interstellar medium, the search for signatures of
antimatter mostly concentrates on antinuclei with $|Z| \ge 2$. 
Although $\overline{He}$ might in principle also be produced in
high energy cosmic ray interactions, their contribution to the
$\overline{H}e^{4}/He^{4}$ ratio is expected to be much smaller
than $10^{-12}$.  From the absence of any candidate event, the
{\sc bess} team \cite{nozaki99} deduced an upper limit on
$\overline{H}e^{4}/He^{4}$ of $10^{-6}$ at rigidities between 1
and 16 GV/$c$.  Similar values have been quoted from a test
flight of the Alpha Magnetic Spectrometer ({\sc ams}) on the
Space Shuttle \cite{ams99}.  These results provide the best
evidence for the Galaxy and nearby Universe being made up solely
of matter.  Future experiments of {\sc bess} or {\sc ams} on the
space station aim at at limit of $\overline{H}e^{4}/He^{4} \le
10^{-8}$ probing the antimatter contents of the Universe to more
than 150 Mpc.

Finally, absolute fluxes of proton, helium, and atmospheric muons
are important also for the derivation of the neutrino oscillation
parameters e.g.\ from Super-Kamiokande \cite{sk98a}, since
the atmospheric neutrino flux is proportional to the
normalization of the dominating \cray\ proton and helium fluxes. 
Presently, new balloon borne experiments are in preparation to
reduce particularly the uncertainties of the muon flux at
different atmospheric depths.  This is of great importance also
for upcoming long-baseline neutrino experiments and is a good
example of the interconnection between cosmic ray and particle
physics.

\section{Extensive Air Showers and High Energy Interactions}

Cosmic ray measurements at energies above some $10^{14}$ eV are
performed by large area air shower experiments.  An extensive air
shower (\eas) is a cascade of particles generated by the
interaction of a single high energy primary cosmic ray nucleus or
nucleon near the top of the atmosphere.  The secondary particles
produced in each collision, mostly charged and neutral pions and
kaons, may either decay or interact with another nucleus, thereby
multiplying the number of particles within an \eas.  After
reaching a maximum in the number of secondary particles, the
shower attenuates as more and more particles fall below the
threshold for further particle production.  A disk of
relativistic particles extended over an area with a diameter of
some tens of metres at $10^{14}$ eV to several kilometres at
$10^{20}$ eV can then be observed at ground.  This magnifying
effect of the earth atmosphere allows to instrument only a very
small portion of the \eas\ area and to still reconstruct the
major properties of the primary particles.  It is a lucky
coincidence that at the energy where direct detection of \cray s
rays becomes impractical, the resulting air showers become big
enough to be easily detectable at ground level.  Due to the
nature of the involved hadronic and electromagnetic interactions
and the different decay properties of particles, an \eas\ has
three components, electromagnetic, muonic, and hadronic. 
Extracting the primary energy and mass from such measurements is
not straightforward and a model must be adopted to relate the
observed \eas\ parameters (total number of electrons, muons,
hadrons, shapes of their lateral density distributions,
reconstructed height of the shower maximum, etc.)  to the
properties of the primary particle \cite{kampert01}.  A large
body of experimental data from heavy-ion collisions studied at
{\sc cern} and Brookhaven and from pp-collisions studied at the
{\sc cern sps} and Fermilab Tevatron is available and has been
used to constrain the phenomenological {\sc qcd}-inspired models
entering \eas\ simulations.  However, \cray\ interactions above
the knee are already beyond the maximum {\sc cms}-energy of the
Tevatron.  Furthermore, the very forward kinematic region being
mostly relevant to the propagation of air showers is basically
uncovered by collider experiments.  Finally, effects of possible
quark-gluon plasma formation may affect \eas\ observables
\cite{ridky00}.  Testing air shower simulations and thereby
hadronic interaction models by means of \eas\ data thus is of
interest for particle and \cray\ physics.

\begin{figure}[t]
\centerline{\epsfig{file=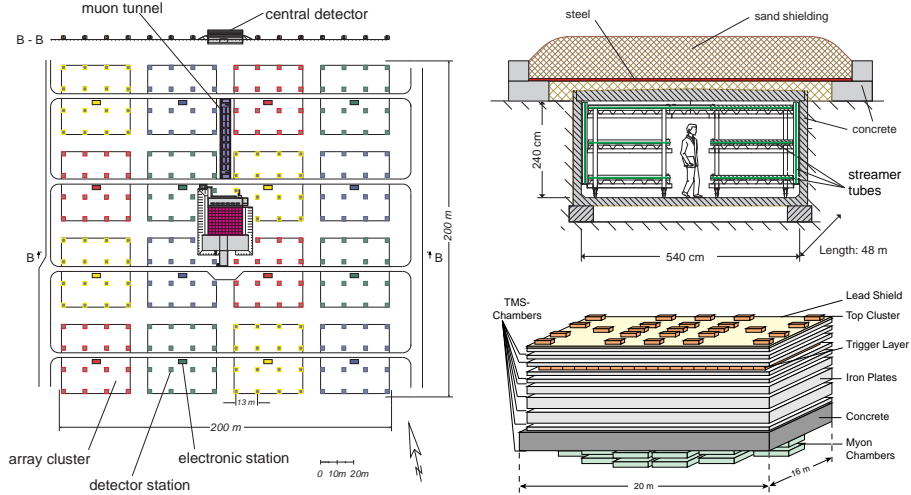,width=12cm}}
\caption[]{Schematic layout of the {\sc kascade} experiment (left), with 
its streamer tube tracking system (top right) and central 
detector (bottom right) \cite{kascade-97c}.}
\label{fig:kascade}
\end{figure}

\begin{figure}[t]
\centerline{\epsfig{file=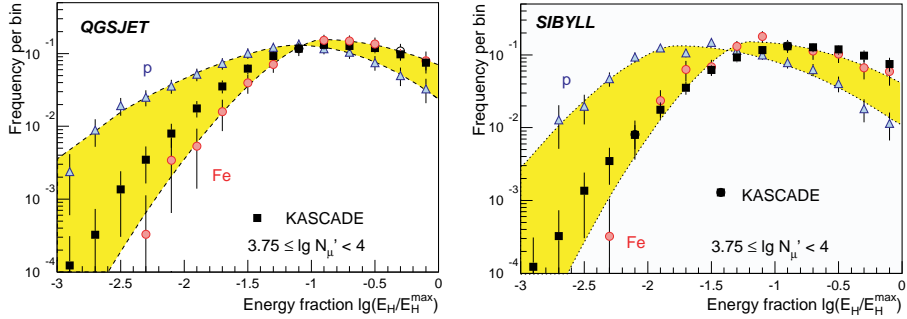,width=12cm}}
\caption[]{Relative energy distribution of hadrons obtained by
normalization to the most energetic hadron in a shower.  The data
are compared to {\sc qgsjet} and {\sc sibyll} simulations for primary protons
(p) and iron nuclei (Fe).  Shaded is the physically meaningful
region as obtained from the simulations.  The primary energy
correspond to 2 PeV. From Ref. \cite{kascade-99c}.}
\label{fig:ww-test}
\end{figure}

The {\sc kascade} experiment at Forschungszentrum Karlsruhe
(Germany), shown in Fig.\,\ref{fig:kascade}, is a $200 \times
200$ m$^{2}$ multi-detector installation measuring all of the
three \eas\ components simultaneously \cite{kascade-97c}.  It
comprises 252 detector stations housing electron-gamma and muon
detectors, a $48 \times 5.4$ m$^{2}$ tunnel for muon tracking,
and a 320 m$^{2}$ large central detector consisting of a finely
segmented hadronic calorimeter (11 $\lambda_{I}$) with additional
muon detection capabilities.  The major goal of the experiment is
to measure the energy spectrum and chemical composition of \cray
s in the energy range of the knee and to allow for tests of the
aforementioned interactions models.  High energy hadrons ($E \ga
100$ GeV) observed at ground by means of the hadronic calorimeter
are easily recognized to provide the best test bench for these
models.  To perform such tests, the {\sc kascade} collaboration
has followed different approaches.  A sensitive test at primary
energies around 10-100 TeV is provided by comparing experimental
and simulated trigger rates in the central detector
\cite{risse99}.  Feeding absolute \cray\ fluxes as measured on
balloons and satellites into the {\sc corsika} air shower
simulation package \cite{corsika} and subsequently into a {\sc
geant}-based detector simulation, allows to directly compare the
expected trigger and hadron rates with experimental data.  This
test exhibits differences between interaction models by about a
factor of two and proves to be sensitive to percentage changes of
the total inelastic cross section or to the contribution of
diffractive dissociation \cite{kascade-01b}.  Another type of
test has been performed by investigating distributions of high
energy hadrons observed in the shower core, an example of which
is given in Fig.\,\ref{fig:ww-test} \cite{kascade-99c}.  Clearly,
the {\sc sibyll} model \cite{sibyll16} provides only a poor
description of the experimental data and has been refined by
know.  At present, {\sc qgsjet} \cite{qgsjet} yields the best
overall result and is used as the `reference model' by most \eas\
experiments.

\section{Mystery of the Knee}

High energy cosmic rays do not only serve as test bench for
hadronic interaction models or as input to atmospheric neutrino
calculations, but are even more interesting in their own right. 
Their origin and acceleration mechanism have been subject to
debate for several decades.  Mainly for reasons of the required
power the dominant acceleration sites are generally believed to
be supernova remnants in the Sedov phase.  Naturally, this leads
to a power law spectrum as is observed experimentally.  Detailed
examination suggests that this process is limited to $E/Z \la
10^{15}$\,eV. Curiously, this coincides well with the knee at
$E_{\rm knee} \cong 4 \cdot 10^{15}$\,eV, indicating that the
feature may be related to the upper limit of acceleration.  The
underlying picture of particle acceleration in magnetic field
irregularities in the vicinity of strong shocks suggests the
maximum energies of different elements to scale with their
rigidity $R=pc/Ze$.  This naturally would lead to an
overabundance of heavy elements above the knee, a prediction to
be proven by experiments.  A change in the \cray\ propagation
with decreasing galactic containment at higher energies has also
been considered.  This rising leakage results in a steepening of
the \cray\ energy spectrum and again would lead to a similar
scaling with the rigidity of particles but would in addition
predict anisotropies in the arrival direction of \cray s with
respect to the galactic plane.  Besides such kind of
`conventional' source and propagation models
\cite{drury94b,berezhko99}, several other hypotheses have been
discussed in the recent literature.  These include an
astrophysically motivated single source model \cite{erlykin97a},
as well as several particle physics motivated scenarios which try
to explain the knee due to different kinds of \cray-interactions. 
For example, photodisintegration at the source \cite{candia00},
interactions with gravitationally bound massive neutrinos
\cite{wigmans00}, or sudden changes in the character of
high-energy hadronic interactions during the development of
extensive air shower (\eas) \cite{nikolsky95} have been
considered.

\begin{figure}[t]
\centerline{\epsfig{file=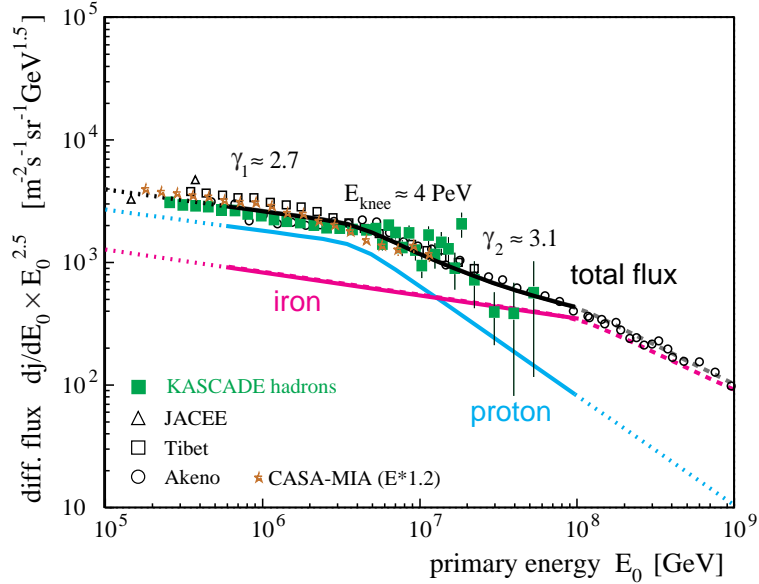,width=10cm}}\vspace*{-2mm}
\caption{Primary energy spectrum in the knee region as obtained
from different experiments.  The lines represent a preliminary
deconvolution of the all-particle {\sc kascade} spectrum into a
light and heavy component \cite{glasstetter99}.}
\label{fig:e-spec}
\end{figure}

To constrain the {\sc sn} acceleration model from the other
proposed mechanisms, precise measurements of the primary energy
spectrum and particularly of the mass composition as
\begin{wrapfigure}[20]{l}{7cm}
\epsfig{file=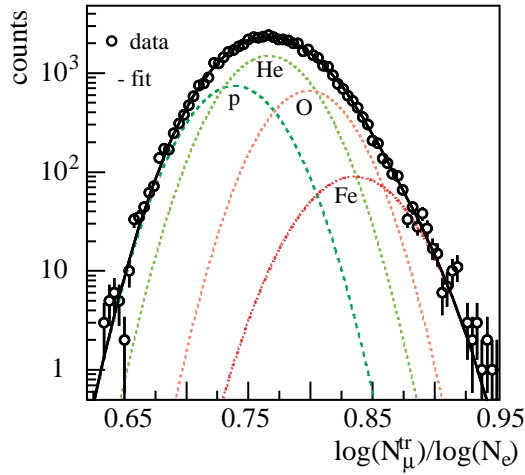,width=7cm}\vspace*{-4mm}
\caption{Muon/Electron ratio from experimental data
compared to simulations using different primary mas-
ses at energies of about 1 PeV \cite{weber99}.}
\label{fig:mu-e-ratio}
\end{wrapfigure} 
a function
of energy are needed.  A number of \eas\ observables has been
identified to serve these purposes with only moderate model
dependence \cite{kampert01}.  Basically, one makes use of the
fact that a heavy primary particle will -- on average --
experience its first interaction higher in the atmosphere as
compared to a proton.  One of the consequences is a stronger
effect of absorption to electrons and photons in the atmosphere
while the more penetrating muons reach ground mostly unaffected. 
Thus, the total number of muons provides a good measure
of the primary energy while the muon/electron ratio is indicative
for its mass.  This is nicely illustrated in
Fig.\,\ref{fig:mu-e-ratio}.  The ratio, calculated on an
event-by-event basis, is compared to {\sc corsika} simulations
using different primary masses.  Interestingly, the entire
elemental spectrum between proton and iron is needed to describe
the experimental data.  Also, the left and right hand tails of
the data are well described by the simulations, giving some
confidence in the reliability of the \eas\ simulations. 
Repeating the analysis in different bins of energy finally yields
the primary energy spectrum for different elemental groups.  The
result of a preliminary analysis is presented in
Fig.\,\ref{fig:e-spec} together with data from experiments other
than {\sc kascade}.  The agreement appears reasonable and
deviations are mostly explained by uncertainties in the energy
scale by up to 25\,\%, e.g.\ {\sc casa mia} data
\cite{glasmacher99a} were shifted upwards in energy by 20\,\% to
yield a better agreement to the other data sets.  This is likely
to be explained by the outdated interaction model {\sc sibyll 1.6}
\cite{sibyll16} employed by the authors of
Ref.\,\cite{glasmacher99a}.  The lines represent fits to the
electron and muon size spectra of {\sc kascade} assuming the
all-particle spectrum to be described by a sum of proton and iron
primaries \cite{glasstetter99}.  Interestingly, a knee is only
reconstructed for the light component and no indication of a
break is seen in the heavy one up to $\sim 10^{17}$~eV. This
important finding giving direct support to the picture of
acceleration in magnetic fields (see above) will be target of
future studies with improved experimental capabilities.  For
example, {\sc kascade} and {\sc eas-top} have just started a
common effort to install the {\sc eas-top} scintillators at the
site of Forschungszentrum Karlsruhe providing a 12 times larger
acceptance as compared to the original {\sc kascade} experiment
and still taking advantage of the multi-detector capabilities.

\section{The Most Energetic Particles in the Universe}

Cosmic rays with energies in excess of $10^{20}$ eV are very rare
(about 1 particle per km$^{2}$ and century) but have been known
for more than 30 years.  There is continuing fascination in
understanding their origin and the route by which they acquire
their macroscopic energy up to 50 Joule or more
\cite{hillas84,nagano00b}.  The Lamor radius of protons or nuclei
at these energies is too large to allow for conventional
acceleration in magnetized shocks within our galaxy.  Searching
the sky beyond our galaxy, hot spots at the termination shock of
gigantic plasma jets streaming out from the central engines of
active galactic nuclei ({\sc agn}) stand out as the most likely
sites from which particles can be hurled at Earth with Joules of
energy.  Sufficiently powerful sources, however, are found only
at distances much larger than 100 Mpc.  This is a major problem
if the \cray\ particles are ordinary nucleons or nuclei, because
soon after the discovery of the cosmic microwave background
radiation it was realized that the universe would be opaque for
protons with $E \ge 6\cdot10^{19}$ eV. This was become known as
the Greisen-Zatsepin-Kuz'min (\gzk) cut-off.  The principal
reaction is $p + \gamma_{2.7K} \to \Delta^{+} \to n + \pi^{+}$ or
$\to p + \pi^{0}$ with a mean free path of $\sim 6$ Mpc. 
Similarly, nuclei of mass $A$ suffer from photodisintegration $A
+ \gamma_{2.7K} \to (A-1) + N, (A-2) + 2N$ occuring via giant
resonances at about the same primary energy.  This is another nice
example of classical nuclear and particle physics processes
giving rise to phenomena in the extreme high-energy Universe. 
Energy spectra of dis-
\begin{wrapfigure}[24]{l}{7.5cm}
\vspace*{-1mm}\epsfig{file=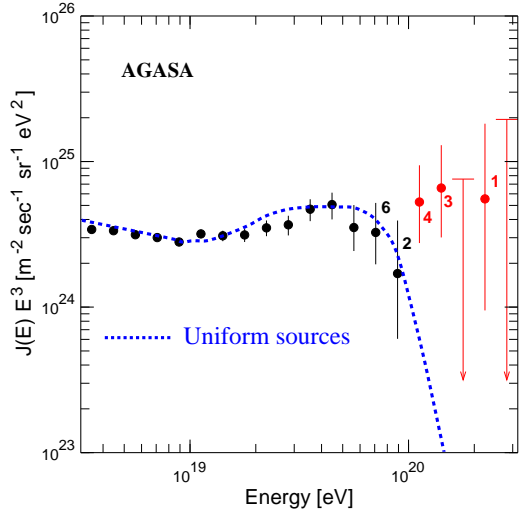,width=7.5cm}\vspace*{-4mm}
\caption{Energy spectrum observed with AGASA \cite{agasa00}.
The vertical axis is multiplied by $ E^{3} $. The dashed curve
represents the spectrum expected for extragalactic sources
distributed uniformly in the Universe, taking account
of the energy determination error.}
\label{fig:agasa-e}
\end{wrapfigure} 
tant sources thus should exhibit a cut-off
as is indicated in Fig.\,\ref{fig:agasa-e}.  However, no such
effect is seen in the experimental data.  Presently, 16 events
with energies above $10^{20}$ eV have been clearly identified
with about half of the statistics originating from the Japanese
{\sc agasa} experiment \cite{agasa00}.  Do we have to conclude from
this that the sources are very close, instead?  If true, one
expects to `see' their sources in the arrival direction of the
\cray s, because known magnetic fields would deflect particles of
such energies by one or two degrees at most.  But again, no
convincing astrophysical source can be identified beside
some doublets and even one triplet of arrival directions.  This
has caused some debate of whether or not we start to see point
sources of extremely energetic \cray s.

The present situation on this fundamental problem appears rather
curious; there is sufficiently convincing experimental material
about the existence of the particles but the statistics is still
too poor to allow for definitive conclusions about their origin! 
An enormous
number of papers and large number of review articles
have addressed the problem of solving the enigma, see e.g.\
\cite{bhattacharjee99,nagano00b,biermann00b}.  Roughly, they can
be grouped into models trying to circumvent the transport
problems of hadrons thereby allowing for distant sources, or novel
exotic sources are invented close by so that the \gzk\ cut-off is
irrelevant.

In the first of the two strategies, primary particles are
proposed whose range is not limited by the {\sc cmb}.  Within the
standard model the only candidate is the neutrino, whereas in
supersymmetric extensions of the standard model, new neutral
hadronic bound states of light gluinos, so called R-hadrons that
are heavier than the nucleon have been suggested.  They would
shift the \gzk\ cut-off to higher energies and thus allow for
$10^{20}$ eV particles.  The particles itself would be produced
as secondaries in collisions of `ordinary' $E \ga 10^{21}$ eV
particles within the powerful ({\sc agn}) accelerator.  Another
proposed solution of the transport problem would be possible
small violations or modifications of fundamental tenets of
physics, e.g.\ violations of Lorentz invariance.  Indeed, such a
violation is expected from models of quantum gravity.  Since all
of these examples require particles to be accelerated to
extremely high energy within the accelerator, this group of
models is often named {\it Bottom Up} approaches.

{\it Top Down} scenarios on the other hand involve the decay of
$X$-particles of mass close to the Grand Unified Theory ({\sc
gut}) scale ($\sim 10^{24}$ eV).  Basically, they can be produced
in two ways: if they are short lived, as expected in many {\sc
gut}'s, they have to be produced continuously.  The only way this
can be achieved is by emission from topological defects (cosmic
strings, magnetic monopoles, domain walls, etc.)  left over from
cosmological phase transitions that may have occurred in the
early Universe at temperatures close to {\sc gut} scale, possibly
during reheating after inflation.  Alternatively, $X$-particles
may have been produced directly in the early Universe.  Due to
unknown symmetries they could have lifetimes comparable to the
age of the Universe.  In all of the pictures, such particles
would contribute to the dark matter and their decays $X \to W,Z
\to q\bar{q}, \gamma, \nu$ would account for the extremely
energetic \cray\ particles.  In this case, the flux would be
dominated by $\gamma$'s and $\nu$'s with only 10-20\,\% nucleons.

Clearly, there is an urgent need to collect a sufficient amount of
data in the \gzk\ energy domain in order to discriminate such
kind of models.  The Pierre Auger Observatory \cite{auger-www},
presently under construction, has been conceived to provide such
data.  The completed observatory will consist of two instruments,
constructed in the northern and southern hemispheres, each
covering an area of 3000 km$^{2}$.  It will be a hybrid detector
system with 1600 particle detectors and 4 eyes of atmospheric
fluorescence telescopes.  The particle detectors will be water
\v{C}herenkov tanks of 10 m$^{2}$ size and 1.2 depth arranged on
a grid of 1.5 km.  During clear moonless nights, the fluorescence
detectors will record the light tracks generated by charged
particles of \eas\ up to distances of 30 km.  This will provide a
very reliable energy (and mass) measurement of the \cray s. 
First data are expected by late 2002.

There are also plans to launch a dedicated satellite or to
instrument a flourescence camera on the {\sc iss} for observing
extremely high energy \cray s from space with much larger
exposure \cite{euso-www} than expected for the Pierre Auger
Observatory.

\section{Summary and Conclusions}

Cosmic ray physics is entering a renaissance of activity and has
become the central pillar of what is known as
Particle-Astrophysics.  The other pillars are TeV
$\gamma$-astronomy addressed by imaging atmospheric \v{C}herenkov
telescopes, TeV-PeV $\nu$-astronomy studied deep underwater or in
ice, and dark matter searches.  It is fascinating to recognize
the Universe as a laboratory for truly high energy physics and to
observe traces of the high energy processes in each of these
different observables.  Due to sophisticated experimental
techniques we are beginning to be able asking much more specific
questions about \cray s, $\nu$s and $\gamma$s at all energies
than have been possible a few years ago and we and may expect to
obtain the answers within the next few years.  A major potential is
given also by synergistic effects combining particle, nuclear,
and atomic physics with astrophysics and cosmology making an
impact on fundamental physic questions.  Examples were given on
how the input from nuclear and particle physics has advanced the
understanding of \cray s and vice versa.  Furthermore, extremely
energetic \cray s address fundamental questions of physics at
energy scales of Grand Unified Theories, i.e.\ at an energy scale
not accessible to man-made particle accelerators.  All of this
constitutes a challenge to basic science and the future appears
very promising and exciting due to the advent of several large
experiments.

\section*{Acknowledgements}
It is a pleasure to contribute to this volume commemorating
Michael Danos.  Mike's interests in science were widely oriented,
he was always open minded and rightfully sceptical.  I sincerely
thank Walter Greiner and co-organizers for inviting me to this
Symposium.  Special thanks go my collaborators in {\sc kascade}
and {\sc auger}.  During the preparation of this talk I
particularly benefited also from fruitful discussions with P.\
Biermann, G.\ Sigl, M. Simon, and many others.

 
\vfill\eject
\end{document}